\documentclass[12pt]{JHEP3}

\usepackage{amssymb}
\usepackage{amsmath}
\usepackage{graphicx}
\usepackage{latexsym}

\newcommand{\psla}{p\kern-.45em/}

\newcommand{\be}{\begin{equation}}
\newcommand{\ee}{\end{equation}}
\newcommand{\beq}{ \begin{eqnarray} }
\newcommand{\eeq}{ \end{eqnarray} }
\newcommand{\beqstar}{ \begin{eqnarray*} }
\newcommand{\eeqstar}{ \end{eqnarray*} }

\newcommand{\GEV}{ {\rm GeV} }
\newcommand{\TEV}{ {\rm TeV} }

\title{\Large Search for the Elusive Higgs Boson Using Jet Structure at LHC}
\author{Chuan-Ren Chen\\
Institute for the Physics and Mathematics of the Universe, \\
The University of Tokyo, Chiba, 277-8583, Japan}
\author{Mihoko M. Nojiri\\
Institute for the Physics and Mathematics of the Universe, \\
The University of Tokyo, Chiba, 277-8583, Japan\\
Theory Group, KEK,1-1 Oho, Tsukuba, Ibaraki 305-0801, Japan\\
The Graduate University for Advanced Studies (Sokendai), \\
1-1 Oho, Tsukuba, Ibaraki 305-0801, Japan}
\author{Warintorn Sreethawong\\
Theory Group, KEK,1-1 Oho, Tsukuba, Ibaraki 305-0801, Japan\\
The Graduate University for Advanced Studies (Sokendai), \\
1-1 Oho, Tsukuba, Ibaraki 305-0801, Japan}

\preprint{
KEK-TH-1369\\
IPMU 10-0091
 }

\abstract{

We consider the production of a light non-standard model 
Higgs boson of order $100~\GEV$ with an associated $W$ boson 
at CERN Large Hadron Collider. We focus on an interesting scenario that, 
the Higgs boson decays predominately into two light scalars $\chi$ 
with mass of few GeV which  sequently decay into four gluons, 
i.e. $h\to 2\chi \to 4g$. Since $\chi$  is much lighter than the Higgs boson, 
it will be highly boosted and its decay products, the two gluons, 
will move close to each other, resulting in a single jet for $\chi$ decay 
in the detector. By using electromagnetic calorimeter-based and 
jet substructure analyses, we show in two cases of different $\chi$ masses 
that it is quite promising to extract the signal of  Higgs boson 
out of large QCD background.

}
\begin{document}
\section{Introduction}

Even though the Standard Model (SM) successfully explains 
almost all of the data in collider experiments, the Higgs boson, 
which is associated with the electroweak symmetry breaking 
mechanism in the SM, has not yet been found. 
The current limits on the Higgs boson mass, $m_h$, come from 
the direct  searches at LEP II and Tevatron, where a  lower bound 
of $114$ GeV has been obtained and a window of 
$158~{\rm GeV} < m_h < 175~{\rm GeV}$ has been excluded, 
respectively, at $95\%$ confidence level (C.L.) ~\cite{LEP, Aaltonen:2010yv}. 
Combined with the direct search limit, the global fits to 
electroweak precision data prefer a light Higgs boson, 
$m_h \lesssim 191~{\rm GeV}$ at $95\%$ C.L.~\cite{LEPEWWG}. 
The Large Hadron Collider (LHC) at CERN in-operation is expected 
to discover the SM Higgs boson with mass up to $1$ TeV, 
if such a particle exists. At the LHC, the Higgs boson is predominately 
produced in  the gluon-gluon fusion process via a top quark loop. 
Other important production processes are vector-boson fusion and 
production of the Higgs boson with an associated $W/Z$ boson 
(associated $hW/hZ$ production). The  search strategy for the Higgs boson depends on  its mass. When the Higgs boson is heavy 
($m_h \gtrsim 135$ GeV), the channels of 
$h\to W W^{(*)}\to \ell_1^+\ell_2^-\nu_1\bar{\nu}_2$ and 
$h\to ZZ\to \ell_1^+\ell_1^-\ell_2^+\ell_2^-$ are very promising. 
For a light Higgs boson ($m_h \lesssim 135$ GeV), 
inclusive $h\to \gamma\gamma$ and $h\to\tau^+\tau^-$ 
in vector-boson fusion production process are the main search channels 
while the dominant decay channel $h\to b\bar{b}$ suffers from 
a very large QCD background.

Recently, the associated $Zh/Wh$ production combined with 
$h\to b\bar b$ search, which  has been considered less promising, 
was recovered by focusing on the regime where the Higgs boson 
is highly boosted and employing a jet substructure technique, and 
it was reported that the significance about $4\sigma$ can be achieved 
for 30 ${\rm fb}^{-1}$ integrated luminosity at the LHC \cite{VH_subjet, subjet_ATLAS}. The idea is that when a heavy particle is highly boosted, 
its decay products will be emitted very collinearly so that they are likely 
to be merged into a single jet rather than to appear as separate jets. 
Such jet should have internal structures corresponding to decay products, 
including their soft/collinear radiations, and the jet mass should peak 
around the mass scale of the parent particle.
Jet substructure has also been applied in searches for 
a boosted top quark decay~\cite{top_tagger,ttH_subjet}, 
strongly interacting $W$ bosons \cite{Seymour} and 
supersymmetric particles \cite{susy_subjet}.

Furthermore, in spite of the great agreement between the SM predictions 
and experimental results, it is widely believed that the SM is only 
an effective theory at the weak scale due to, for instance,  the so-called 
"hierarchy problem" and lack of a dark matter candidate in the SM. 
Many new models beyond the SM have been proposed over past decades, 
and in most of them the Higgs sector is extended. It should be emphasized 
that the current bound on the Higgs boson mass cannot be directly applied 
to such models especially when properties of the Higgs sector are significantly modified. For example, the LEP lower bound is based on studies of 
the associated $Zh$ production with $h\to b\bar b$. Therefore, 
the limit can be relaxed if the production cross section, or equivalently 
an effective coupling $g_{ZZh}$, is suppressed. Another interesting possibility 
is that the Higgs boson is hidden, i.e. it has non-SM decay modes 
and escapes from detection even though it might have been produced 
at LEP and Tevatron. Particularly, if the decay of the Higgs boson into 
not-yet-discovered light particles dominates, the direct search strategies 
have to be revised to capture the nature of new decay topologies. 
Examples of scenario are the extensions of Minimal Supersymmetric Standard Model (MSSM) including an additional singlet \cite{NMSSM, NMSSM_LHC} 
or CP-violating interactions \cite{CPX}, the little Higgs models \cite{LH} and supersymmetric little Higgs models \cite{susyLH_gluon, susyLH_charm}. 
In most of these models, the Higgs boson dominantly decays into 
a pair of new light scalars $\chi$ which subsequently decay into two 
or more visible SM particles. The phenomenology of such non-SM decays 
has been studied both at the Tevatron \cite{pheno_TEV, pheno_TEV_LHC} 
and at the LHC \cite{NMSSM_LHC, pheno_TEV_LHC, pheno_LHC}. 
The most commonly studied cascade decays are 
$h\to 2\chi\to 4b/4\tau/2b2\tau$.

In this paper, we study a scenario in which  a light Higgs boson 
($\lesssim 135$ GeV) decays predominately into a pair of very light 
pseudoscalars, $\chi$, ( $m_\chi < 2m_b$)
each of which then decays into two gluons, i.e. $h\to2\chi\to 4g$. 
A model which predicts such a decay channel will be briefly described 
later in Section~\ref{sec:model}. However, we stress that we  do not restrict ourselves to a particular model since we treat the branching ratios of 
$h\to 2\chi$ and $\chi\to 2g$ as free  parameters. 
This channel will encounter the large QCD background, we therefore 
focus on the associated $Wh$ production process where $W$ boson 
decays leptonically in order to suppress the background. 
Since the  mass splitting between Higgs boson and $\chi$ particle is large, 
two pseudoscalars will be highly boosted. An opening angle between 
two gluons from each $\chi$ decay, which is also limited by 
the small mass of $\chi$ will be so small that they are closely aligned 
and typically present in a single jet  in the detector. The collider signature is essentially two central non-$b$ jets, an isolated charged lepton and missing momentum. We will show that the signal for a relatively light $\chi$
can be feasibly discovered at the LHC by imposing cuts on electromagnetic-calorimeter-based (E-cal-based) variables, together with some kinematic cuts. 
For the case of heavier $\chi$ the angular separation between 
daughter gluons from a pseudoscalar decay is larger, so 
the E-cal-based analysis fails. However, by employing the jet substructure 
technique previously mentioned, we demonstrate that such a scenario 
still can be discovered. For illustration, we adopt $m_h=120~\GEV$, 
$100\%$ decay branching ratio for $h\to2\chi$ and $\chi\to 2g$, 
and two masses for $\chi$: $m_\chi = 4$ GeV (light) and 
$m_\chi= 8$ GeV (heavy).

The paper is organized as follows. In the next section, we firstly give 
a brief review on the model of elusive Higgs boson. Afterward, signal and
background generations are described. In Section 3, we present 
a search strategy using narrow jet at the LHC for the case $m_\chi = 4~\GEV$. 
Jet substructure method and its application to the case $m_\chi=8~\GEV$ 
are presented in Section 4. Lastly, Section 5 is devoted to discussions and conclusions.

\section{Higgs Decay $h\to 2\chi\to 4g$ at the LHC}

\subsection{A Reference Model}
\label{sec:model}
For  models beyond the SM, it is not unusual that a light singlet scalar appears as a result in the extended Higgs sector. If there exists a coupling between the Higgs boson and the light scalar, the Higgs may predominately decay into a pair of such light scalars. The dominant decay channel of the light scalar then depends on its mass and couplings to SM fermions. Here we briefly review a supersymmetric extension of the simplest Little Higgs model~\cite{susyLH_gluon},  in which the light scalar will mainly decay into a gluon pair when its mass is lighter than twice the bottom quark mass, which is the scenario we focus on in this paper.
We only introduce the phenomenology in Higgs sector and we refer readers to the original paper for the details of the model.  

The model is a supersymmetric extension of the simplest Little Higgs model~\cite{susyLH_gluon}.  The gauge group is $SU(3)_C\times SU(3)_W\times U(1)_X$, where the $SU(3)_W\times U(1)_X$ is broken by two vector-like sets of Higgs superfields, $\Phi_{u,d}$ and ${\cal H}_{u,d}$. The misalignment between the vacuum expectation values (VEVs) of these two Higgs fields leads to electroweak symmetry breaking.
There exists an approximate $SU(3)_1\times SU(3)_2$ global symmetry due to the absence of the cross terms between $\Phi$ and ${\cal H}$.  The VEV of $\Phi$ at the order of 10 TeV breaks the $SU(3)_1$ and the gauge group down to the SM  $SU(3)_C\times SU(2)_W\times U(1)_Y$, generating five massive gauge bosons. 
The VEV of ${\cal H}$, which is generated radiatively with an order of  few hundreds GeV, breaks another set of $SU(3)\times U(1) \to SU(2)\times U(1)$ and produces five Goldstone bosons, three of which are eaten by $W$ and $Z$ bosons after electroweak symmetry breaking, and the two remnants are identified as Higgs boson $h$ and light pseudoscalar $\chi$. The coupling between $h$ and $\chi$
is generated due to  the fact that $h$ lives partly in the third component of 
${\cal H}_{u,d}$, where $\chi$ mainly stays. 
After the Higgs gets a VEV and canonically normalizing the Higgs and pseudoscalar fields, the tree-level interaction between $h$ and $\chi$ generated from the Higgs kinetic terms is given as~\cite{susyLH_gluon}
\begin{eqnarray}
{\cal L}_{h\chi\chi} \approx -\frac{h}{\sqrt{2}v_{\mathcal H}}(\partial_\mu \chi)^2\frac{v_{em}}{\sqrt{v_{\mathcal H}^2 - v_{em}^2}}.
 \end{eqnarray} 
 The coupling of the Higgs boson to the SM fermions is the same as that in the SM but with an additional factor $\sqrt{1-v_{em}^2/v_{\cal H}^2}$.

The decay widths of the Higgs boson into a pair of $\chi$'s and the SM fermions $f$ are therefore
\begin{eqnarray}
\Gamma_{h\to\chi\chi} &\approx& \frac{1}{64\pi}\left(1-\frac{v^2_{ew}}{v_{\mathcal H}^2} \right)^{-1} \frac{m_h^3 v^2_{ew}}{v_{\mathcal H}^4},\\
\Gamma_{h\to f\bar{f}} &\approx& \left(1-\frac{v^2_{ew}}{v_{\mathcal H}^2} \right)\Gamma^{\rm SM}_{h\to f\bar{f}},
\end{eqnarray}
where $v_{ew}= 174$ GeV is the electroweak scale related to the Higgs VEV, $v_{\mathcal H}\approx 300~\GEV\sim 500~\GEV$ and $\Gamma^{\rm SM}_{h\to f\bar{f}}$ is the decay width of the SM Higgs boson into a fermion pair. 
After electroweak symmetry breaking, $\chi$ acquires mass naturally of order few to ${\cal O}(10)$ GeV. The coupling between the singlet $\chi$ 
to the SM fermion $f$ is induced by the mixing of the SM fermion and its heavy partner $f'$, and is suppressed by a factor of $m^2_{f}/m^2_{f'}$. 
Therefore, only the couplings of third generation fermions to 
$\chi$ should be considered.
When $\chi$ is heavier than $2m_b$, it decays almost $100\%$ into a pair of $b$ quarks. Below the threshold of two $b$ quarks, the dominant decay channel of 
 $\chi$ is always $\chi\to g g $ and  the subdominant channel (branching ratio $< 1\%$) is $\chi\to\gamma\gamma$ or $\chi\to\tau^+\tau^-$ depending on the mass of $\chi$. For example, with the parameters chosen in the Fig. 7 of the Ref~\cite{susyLH_gluon} the  branching ratio of $\chi\to\gamma\gamma$ is larger (smaller) than  $\chi\to\tau^+\tau^-$ when $\chi$ is heavier (lighter) than about $8$ GeV. 

In summary, we review in this section a reference model in which the light Higgs boson mainly decays into two singlet light scalars which sequently decay into four gluons, i.e. $h\to 2\chi \to 4g$. 
In our numerical study, we take $100\%$ branching ratio for both 
$h\to 2\chi$ and $\chi \to 2g$ for $m_h = 120~ \GEV$ and $m_\chi < 2m_b$, 
since we only focus on this non-SM Higgs decay scenario and will explore it in a model independent approach. For any particular model which predicts  such a scenario, our result can be easily applied by rescaling the production cross section of the Higgs boson and branching ratios.

\subsection{Signal and Background Generations}

For signal event generation, we employed the Herwig 6.5 Monte Carlo event generator \cite{Herwig} at the LHC with $\sqrt{s}= 14~\TEV$ center-of-mass energy. We generated the heavier neutral MSSM Higgs boson in association with a $W$ boson and forced the Higgs boson to decay into two pseudoscalars each of which subsequently decays into two gluons. 
All backgrounds were generated by Alpgen 2.13 \cite{Alpgen} with showering and hadronization by Herwig. 
In our analysis, we particularly forced $W\to e\nu$ for both signal and background processes, therefore the final statistics should  be twice our result  when $W\to\mu\nu$ is also considered. 
The leading order cross sections incorporating branching fraction for signal and background processes are listed in Table \ref{sigma_ecal}. Note that the signal cross section is normalized to the SM value.
We present in detail two Higgs search strategies for 
$m_\chi = 4~\GEV$ and $m_\chi =8~\GEV$ 
in the next two sections: the electromagnetic calorimeter-based method and the jet substructure method, respectively. For the former analysis, $Wjj$ is the dominant background process. For the jet substructure analysis, $t\bar t, tW$ and $tbW$ backgrounds are also important.

\TABLE[!ht]{
\begin{tabular}{|l|r|}
\hline
process & cross section @ LO (pb)\\
\hline
$hW\to e\nu jj$ & 0.28 \\ 
\hline
$Wjj\to e\nu jj$ & 1401.64 \\
$WW\to e\nu jj$ & 11.16 \\
$t\bar t\to e\nu b\bar b jj$ & 65.53 \\
$tq\to e\nu b j$ & 10.61 \\
$tW\to e\nu b jj$ & 7.99 \\
$tbW\to e\nu bb jj$ & 12.17 \\
\hline
\end{tabular}
\caption{Production cross sections for signal and background processes.}
\label{sigma_ecal}
}

After event generation, we employed  Fastjet \cite{Fastjet} for jet reconstruction.
We used Cambridge-Aachen (CA) and KT jet algorithms in this study. Moreover, all events are subjected to the following pre-selection cuts:\\
\hspace{5pt}(i) an isolated lepton with transverse momentum $p_{Tl}>20~\GEV$,\\
(ii) at least two jets with transverse momenta $p_{T1}> 40~\GEV, p_{T2}> 30~\GEV$ 
and pseudo-rapidities $|\eta_{1,2}|<2$,
\\
(iii) a transverse mass $m_T=\sqrt{2(E_{Tl}E_T^{\rm miss}-{\bf p}_{Tl}\cdot{\bf p}_T^{\rm miss})}<m_W$ (lepton mass is neglected) 
where $E_T^{\rm miss}$ and ${\bf p}_T^{\rm miss}$ are transverse missing energy and momentum respectively,\\
(iv) no tagged $b$-jets.\\
Note that the energy smearing effects of isolated photons, charged leptons and jets are taken into account in the pre-selection cut~\cite{acerdet}.
The cut on transverse mass is used to reduce backgrounds with an accidental isolated lepton which is not from $W$  boson decay. Moreover, we veto $b$-jet to reduce backgrounds from top quark events and assume 60\% $b$-tagging efficiency.

\section{Narrow Jet and $m_\chi=4~\GEV$ Case}

Since $m_\chi = 4~\GEV$ is much lighter than the Higgs boson ($m_h=120~\GEV$), the decay products of $\chi$ will be very collinear, generating a very narrow single jet in the detector. We therefore resolve the narrow jet structure with the aid  of fine granularity of electromagnetic calorimeter, which has been well studied for the hadronic decay of $\tau$ lepton \cite{CERN-OPEN-2008-020}.

We now introduce variables adapted in the reconstruction technique for the hadronically decaying $\tau$, 
which will be used to characterize a jet from $\chi$ decay. The starting point of our analysis is to define jet axes. We use true parton momentum axes of 
$\chi$ particles for signal events, while take two hardest jet axes from jet-clustering algorithm for background events. Then, we define a cone of radius $R=0.4$ around each jet axis and calculate the following quantities:

\begin{itemize}
\item
{\bf The electromagnetic transverse energy $(E^{\rm em}_{T})$}\\
The electromagnetic transverse energy is defined as a sum of transverse energy deposited in the electromagnetic calorimeter and transverse energy of charged particles within a cone radius $R$, i.e.
\beq
	E^{\rm em}_{T}\equiv{\sum_{R<0.4} E^{\rm em}_{T,i}},
\eeq
where $i$ runs over all electromagnetic calorimeter cells and charged particles within a cone of radius $R=0.4$. In this analysis, we take the granularity of electromagnetic calorimeter cells $\Delta\eta\times\Delta\phi=0.025\times 0.025$ 
corresponding to the middle layer of ATLAS calorimeter \cite{CERN-OPEN-2008-020}. 
This variable is similar to the jet transverse energy, however, cut on this variable is tighter and more effective to reduce backgrounds. 

\item
{\bf The electromagnetic radius ($R_{\rm em}$)}\\
The electromagnetic radius ($R_{\rm em}$) is defined as
\beq
	R_{\rm em}=\frac{\sum_i E^{\rm em}_{T,i}
	\sqrt{(\eta_i-\eta_{\rm jet})^2+(\phi_i-\phi_{\rm jet})^2}}{E^{\rm em}_{T}},
\eeq
where ($\eta_i, \phi_i$) and ($\eta_{\rm jet}, \phi_{\rm jet}$) denote (pseudo-rapidity, azimutual angle) of the $i$th cell and the jet axis, respectively. This variable gives an effective cone radius in which a large fraction of energy is deposited. For collimated jets, $R_{\rm em}$ tends to be narrow and peak at small value. However, it should be noted that the efficiency of cut on this variable is $E_T$-dependent. When $E_T$ is increasing, the $R_{\rm em}$ distribution will be narrower. 
Therefore, this variable becomes less effective when $E_T$ is large. 

\item
{\bf The energy isolation in the calorimeter ($E_{\rm iso}$)}\\
The energy isolation is defined as 
\beq
	E_{\rm iso}=\sum_{r_1<R<r_2}E^{\rm em}_{T,i}/E^{\rm em}_{T}
\eeq
where $i$ runs over all electromagnetic calorimeter cells and charged particles inside a ring of $r_1<R<r_2$.
In our study we take $r_1 = 0.1$ and $r_2 = 0.4$.
For highly collimated jets, most QCD activities distribute near the axis of core, i.e. within $R<0.1$, and the $E_{\rm iso}$ distribution is narrow with a peak at zero. 
 Similar to $R_{\rm em}$, performance of $E_{\rm iso}$ cut is also $E_T$-dependent. In addition, this variable is expected to be less effective for events with higher hadronic activity such as $t\bar t$ events.
\end{itemize}

\FIGURE[!ht]{
\hskip -15pt
\includegraphics[scale=0.6]{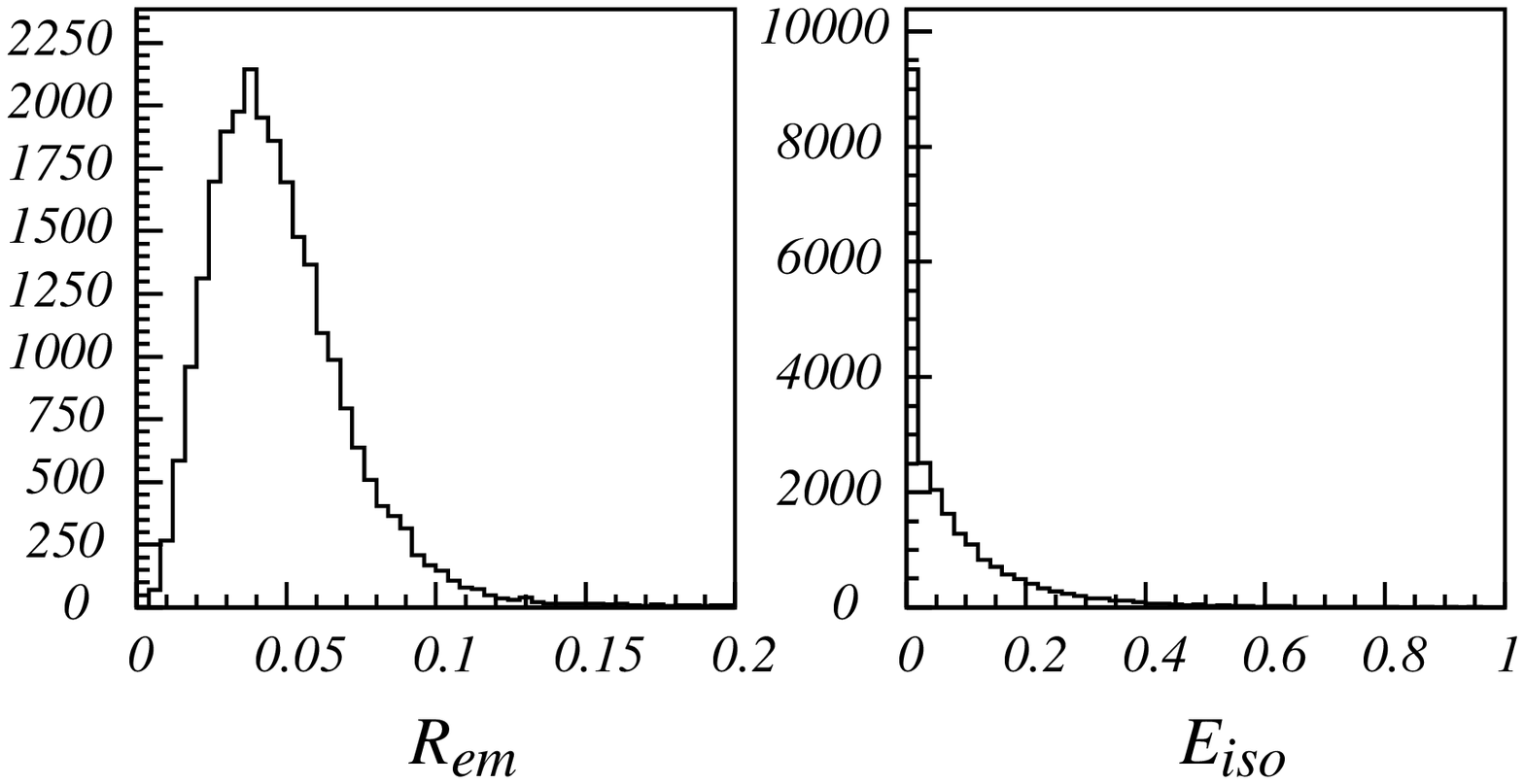}

\includegraphics[scale=0.6]{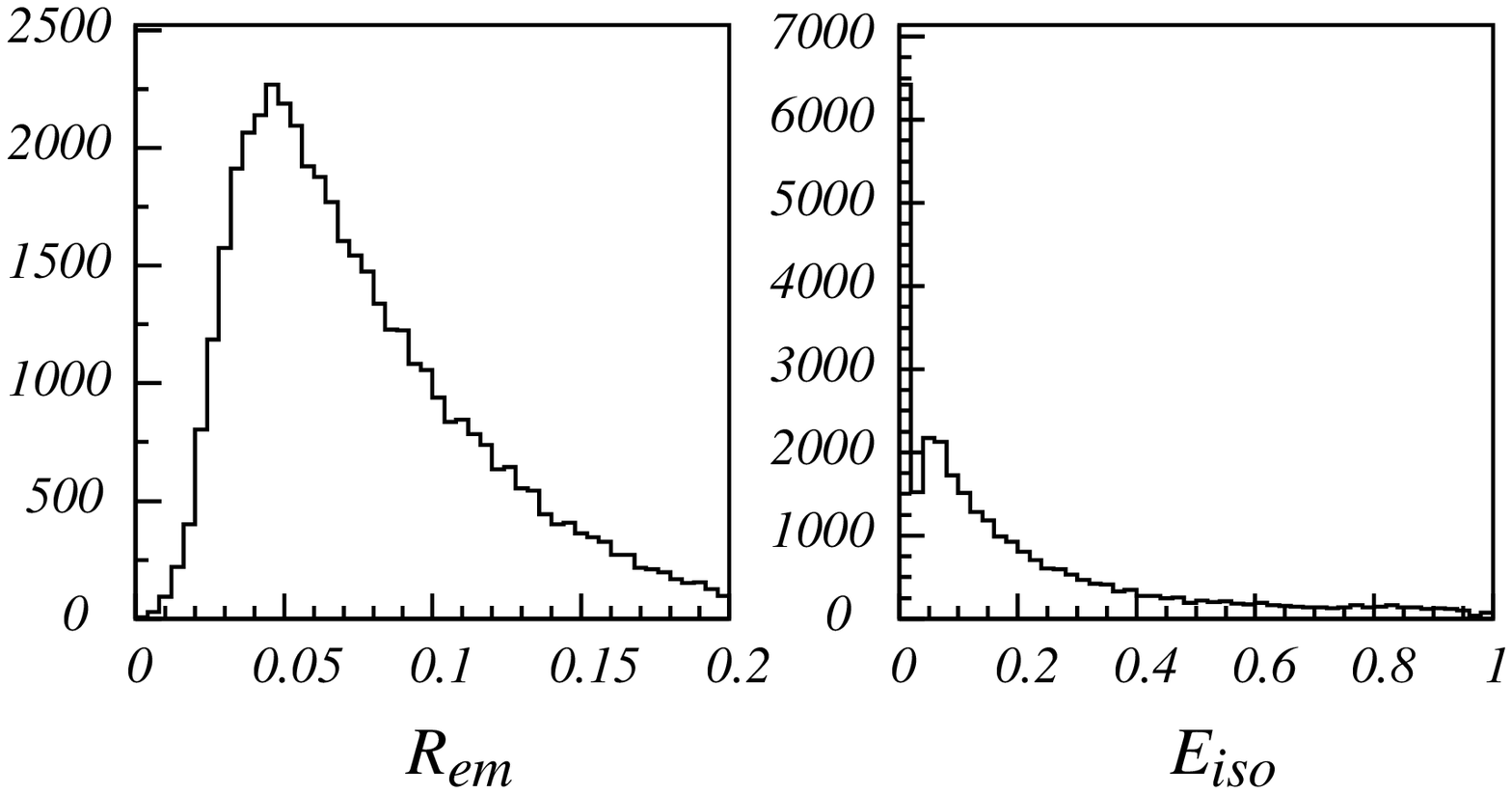}
\caption{ Non-normalized distributions of electromagnetic calorimeter-based variables, $R_{\rm em}$ and $E_{\rm iso}~(E^{\rm em}_{T}>30~\GEV)$, for signal with
 $m_\chi=4~\GEV$ (upper panels) 
 and for $Wjj$ background (lower panels).}
\label{lightHiggs_ecal}
}

We show in Figure \ref{lightHiggs_ecal} the non-normalized distributions of $R_{\rm em}$ and $E_{\rm iso}~(E^{\rm em}_{T}>30~\GEV)$ for $m_\chi =4~\GEV$ signal (upper panels) and for the dominant $Wjj$ background (lower panels).
For comparison, we also show similar distributions for different SM Higgs decay channels: $h\to gg$, $h\to b\bar b$, $h\to u\bar u$, and $h\to\tau^+\tau^-$ in Appendix A. From Figure \ref{lightHiggs_ecal}, the distributions for signal are clearly narrower than those for $Wjj$ background. Almost all signal events has $R_{\rm em}$ smaller than 0.1, while the distribution for $Wjj$ events extends beyond 0.2. Similarly, $E_{\rm iso}$ for signal is less than 0.5, but the background distribution has a long tail up to $E_{\rm iso}\sim 1$. This indicates that high energy objects inside a jet are more highly collimated for signal. 

In our analysis, we choose the following cuts:
\begin{itemize}
\item $R_{\rm em} < 0.06$,
\item $E_{\rm iso} < 0.15$,
\item $E^{\rm em}_{T}(j_1)\ge 100~\GEV$, $E^{\rm em}_{T}(j_2)\ge 50~\GEV$.
\end{itemize}

\FIGURE[!ht]{
\includegraphics[scale=0.6]{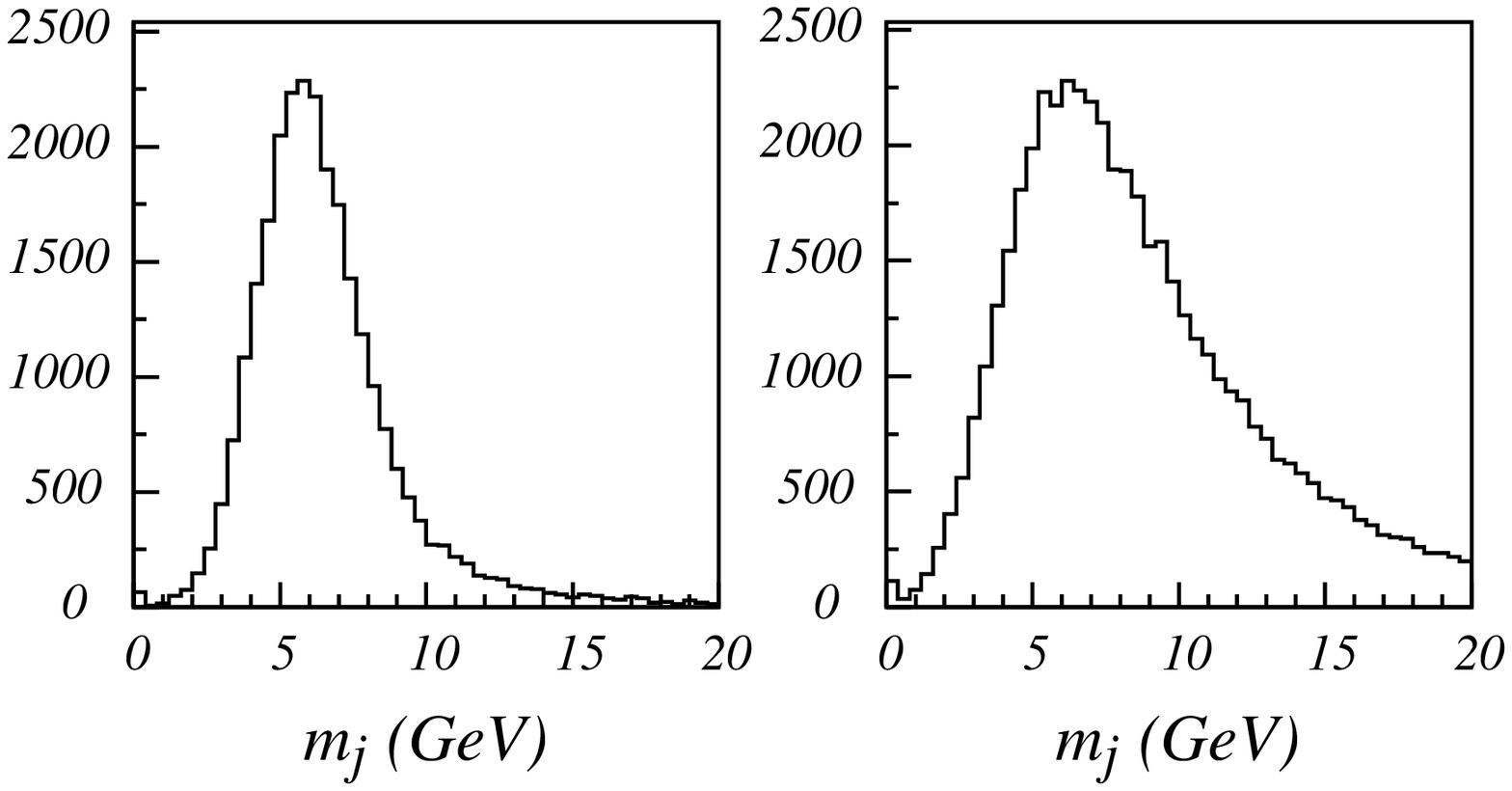}
\caption{Non-normalized jet mass distributions of the first two hardest jets
for signal with $m_\chi=4~\GEV$ (left) 
and for $Wjj$ background (right) after imposing pre-selection cuts.}
\label{jetmass_kt}
}

\TABLE[!ht]{
\begin{tabular}{|l|r|r|r|r|r|r|r|}
\hline
Selection cut & $Wh$ & $Wjj$ & $WW$ & $t\bar t$ & $tq$ & $tW$ & $tbW$ \\
{} & ($m_\chi=4~\GEV$) & & & & & & \\
\hline\hline
pre-selection cuts & 811 & 382,685 & 1,868 & 2,460 & 412 & 644 & 400 \\
\hline
$n_j\le 4$ & 772 & 345,211 & 1,444 & 600 & 380 & 410 & 112 \\
$m_j< 8~\GEV$& 601 & 115,214 & 620 & 288 & 118 & 112 & 52 \\
$R_{\rm em}< 0.06$ & 456 & 37,735 & 194 & 108 & 16 & 40 & 32 \\
$E_{\rm iso}<0.15, E^{\rm em}_{T}>30~\GEV$ & 450 & 25,646 & 125 & 124 & 14 & 32 & 24 \\
$E^{\rm em}_{T}(j_{1,2})\ge 100,50~\GEV$ & 98 & 1,431 & 3 & 56 & 2 & 8 & 0 \\
\hline
after imposing all cuts above& 29 & 17 & 0 & 0 & 0 & 0 & 0 \\
\hline
\end{tabular}
\caption{Number of expected signal and background events after cuts in the dijet invariant mass window 110~GeV $\le m_{jj}\le$ 130~GeV for $\mathcal{L}=30~{\rm fb}^{-1}$ at the LHC. The analysis is based on the KT algorithm.}
\label{ecal_count}
}

In addition to electromagnetic calorimeter-based variables introduced above, we also employ cuts on another two variables: 
the number of jets 
\footnote{The transverse momentum and rapidity of jets defined through this paper are: $p_T>10~\GEV$ and $|\eta| <5$.}

and jet mass. The signal events are expected to contain two jets with few additional jets from initial/final state radiations.
In contrast, $t\bar t$ and $tbW$ backgrounds are expected to have high jet multiplicity in the final states and then can be suppressed by imposing the upper limit on number of jets. The key difference in jet mass distribution between signal and QCD background is the following. For a heavy particle decaying hadronically into a single jet, the jet mass distribution will exhibit an enhancement structure corresponding to the mass scale of the particle. For QCD backgrounds, even though original partons are massless, the QCD splitting can generate nonzero masses to the jets. In perturbative theory, jet mass appears at next-to-leading order (NLO) where two massless partons can be present in a single jet. The average value is approximately given by~\cite{Ellis:2007ib}
\beq
	\langle m^2_j\rangle\simeq C\frac{\alpha_s}{\pi}p^2_TR^2,
\eeq
where $C$ is a coefficient that depends on the type of partons and $R$ is a parameter equivalent to cone radius. At the NLO, there is no difference between two jet recombination algorithms because there is only one way to combine two parton into a jet. Beyond the average jet mass, the distribution falls smoothly due to the lack of any intrinsic mass scale. In Figure \ref{jetmass_kt}, we show non-normalized jet mass distributions from KT algorithm with R=0.4 for signal (left) and $Wjj$ background (right) after imposing pre-selection cuts. The peak of signal distribution is a bit higher than the true pseudoscalar mass due to gluon radiation contribution. Although the distribution for $Wjj$ background seems to peak near the pseudoscalar mass value, however, we should keep in mind that the position of the peak is approximately linear in $p_T$. 
Imposing high-$p_T$ cut along with the upper bound of  jet mass  can be useful for background reduction.

In Table \ref{ecal_count}, we show the number of expected signal and background events in a dijet invariant mass window 110~GeV$\le m_{jj}\le$ 130~GeV 
for $30~{\rm fb}^{-1}$  integrated luminosity at the LHC 
{\bf\footnote{
The momentum smearing is parametrized by Gaussian resolution $50\%\sqrt{E}$ \cite{acerdet}. Therefore, the dijet invariant mass resolution for $p_{Tj_{1,2}}= 100,50~\GEV$ is about 10 GeV.
}.}
The first row shows number when only pre-selection cuts are applied. Lower entries are numbers when cuts on number of jets, jet mass, and electromagnetic calorimeter-based variables are separately imposed. The final row shows the number of events when we use all cuts.

\FIGURE[!ht]{
\hskip -15pt
\includegraphics[scale=0.35]{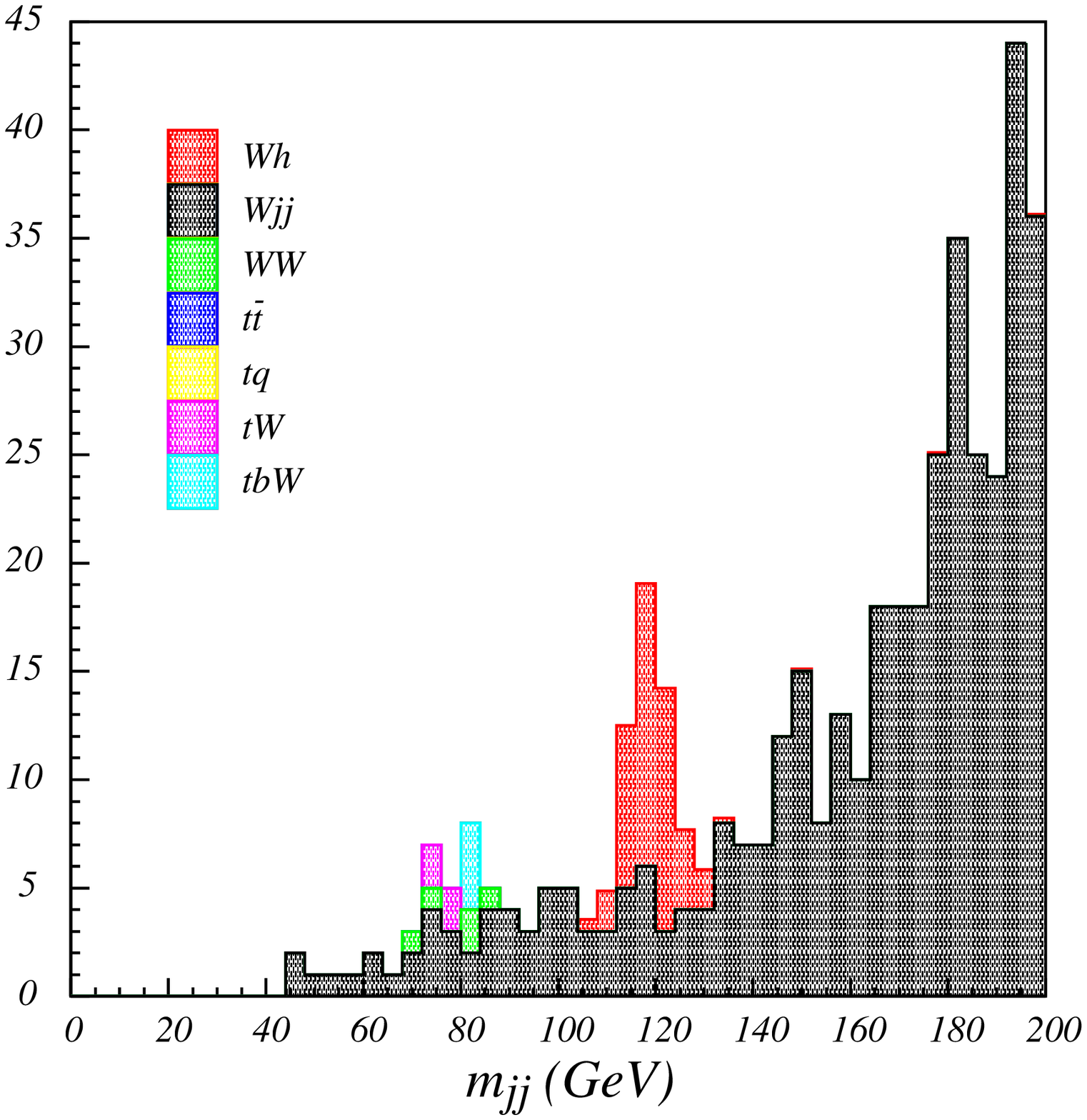}
\includegraphics[scale=0.35]{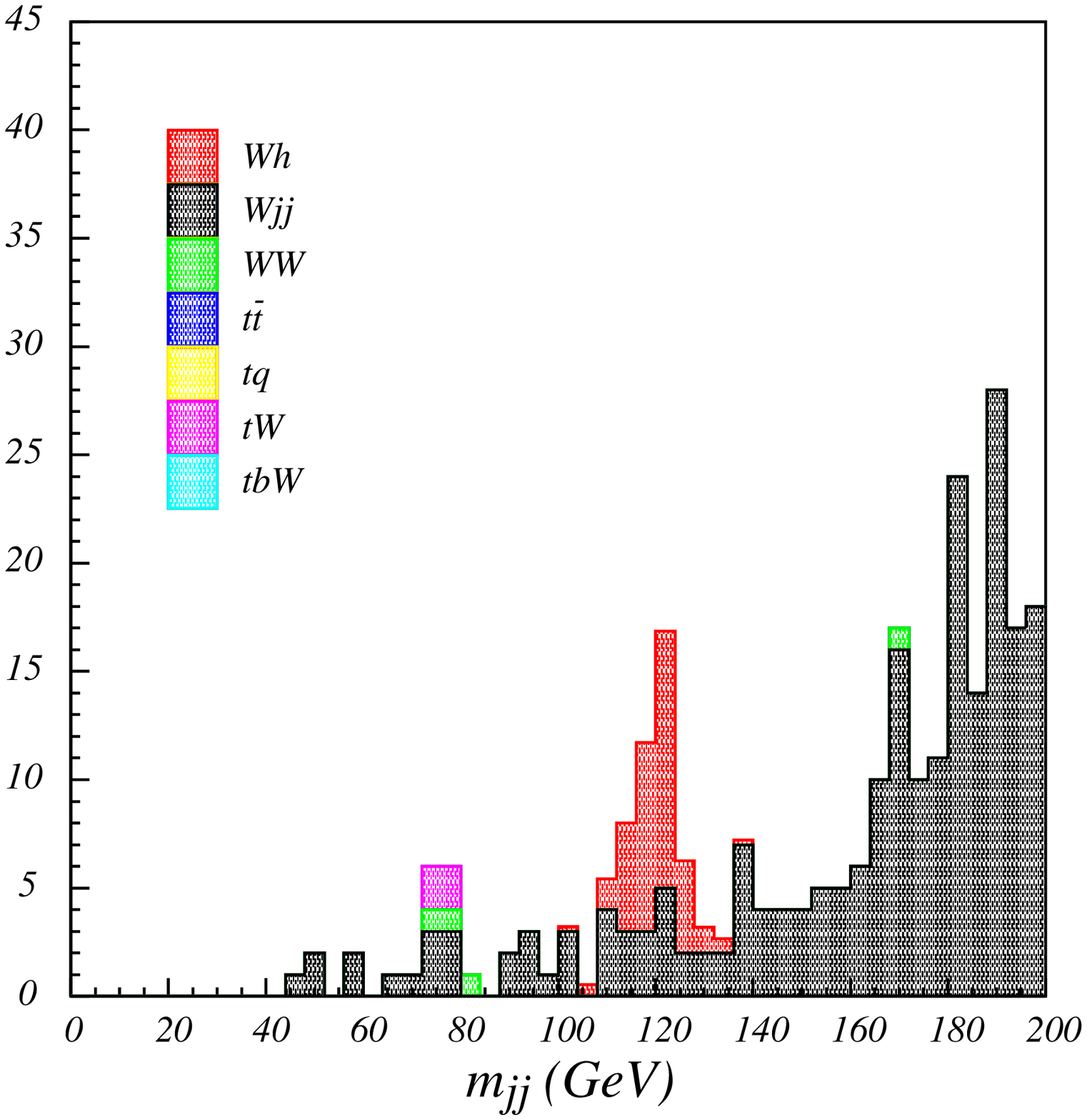}
\caption{Dijet invariant mass distributions for $pp\to Wh$ and
 $h\to 2\chi\to 2j$ $(m_\chi=4~\GEV)$ 
 and SM backgrounds from electromagnetic calorimeter-based analysis using CA algorithm (left) and KT algorithm (right) for $\mathcal{L}=30~{\rm fb}^{-1}$ at the LHC.}
\label{lecal_combine}
}

As we can see in Table  \ref{ecal_count}, after imposing pre-selection cuts, $Wjj$ is the dominant background, and $WW$ and $t\bar t$ with $b$-jets misidentified as leptons are also significant. When $n_j\le 4$ is imposed, $t\bar t$ and $tbW$ events are reduced by a factor of four. By demanding $m_j< 8~\GEV$, the number of backgrounds is reduced by a factor of $3\sim10$. Again, it should be emphasized that these numbers are before applying high-$p_T$ cut on jets.
The efficiency of $R_{\rm em}$ and $E_{\rm iso}$ cuts are impressive. They remove backgrounds by one order of magnitude while keeping more than half of signal events. Lastly, since $E^{\rm em}_T$ cut is equivalent to $p_T$ cut and even tighter, considering high-$E^{\rm em}_T$ events hence corresponds to focusing on a boosted Higgs boson regime for signal. This cut is proved to be very helpful for Higgs discovery in our study. This is because jets in several background processes are coming from hadronic $W$ decay or mistagged $b$-jet. In the case that both jets are from single $W$, it is unlikely that they are simultaneously hard. The condition of $E^{\rm em}_{T}(j_{1,2})\ge 100,50~\GEV$ will select roughly $10\%$ of signal events out of those after imposing pre-selection cuts. While only $0.37\%$ and $2.28\%$ of $Wjj$ and $t\bar t$ events, respectively, can pass this cut, which are, however, still large enough to dominate over signal. On the contrary, all other backgrounds are suppressed to the negligible level. Another advantage of $E^{\rm em}_T$ cut is that the invariant mass of uncorrelated dijet moves toward higher mass region when a harder $E^{\rm em}_T$ cut is adopted, but the signal always peaks at the value of the Higgs boson mass.

\TABLE[!ht]{
\begin{tabular}{|l|c|c|}
\hline
Jet algorithm & $\sigma_S$ (fb) & $S/\sqrt{B}$\\
\hline
CA & 1.13 & 7.09 \\
KT & 0.97 & 7.03 \\
\hline
\end{tabular}
\caption{Signal cross section and statistical significance after all cuts in the dijet invariant mass window 110~GeV $\le m_{jj}\le$ 130~GeV for $\mathcal{L}=30~{\rm fb}^{-1}$ at the LHC.}
\label{ecal_sigma}
}

After applying all cuts altogether, the dijet invariant mass distributions for 
$m_\chi=4~\GEV$ signal and backgrounds are shown in Figure \ref{lecal_combine} for $\mathcal{L}=30~{\rm fb}^{-1}$ at the LHC. The CA (KT) algorithm is used in analysis in the left (right) plot. The Higgs boson signal can be clearly visible above the backgrounds and the performances of CA and KT algorithms  agree well with each other. The signal cross section and statistical significance after imposing all cuts in the dijet invariant mass window 110~GeV $\le m_{jj}\le$ 130~GeV are listed in Table \ref{ecal_sigma}, which shows that  a $7\sigma$ significance level can be achieved.

\section{Jet Substructure and $m_\chi=8~\GEV$ Case}
\label{sec:subjet}

When the mass of $\chi$ becomes larger, the angular splitting between two gluons from $\chi$ decay increases and, on average, the transverse energy spreads into broader region inside a defined cone. This can be clearly seen in the $R_{\rm em}$ and $E_{\rm iso}$ distributions shown in Figure \ref{heavyHiggs_ecal}. They are even harder and broader than those for $Wjj$ background (see lower panels of Figure \ref{lightHiggs_ecal}). As a result, we found that the electromagnetic calorimeter-based method fails as a strategy for Higgs discovery if pseudoscalar particle is not light enough.

\FIGURE[!ht]{
\includegraphics[scale=0.6]{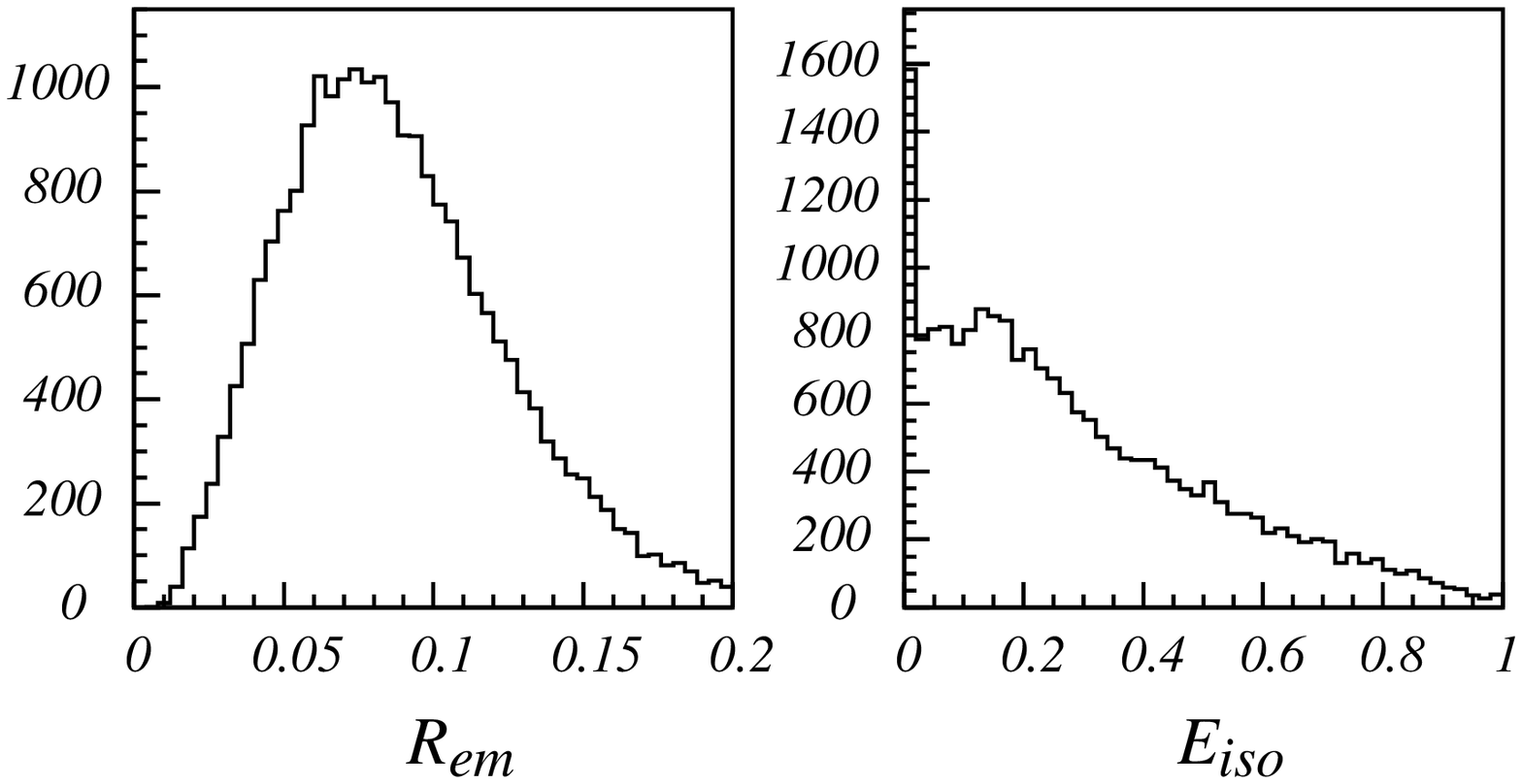}
\caption{Non-normalized distributions of electromagnetic calorimeter-based variables, $R_{\rm em}$ and $E_{\rm iso}$, for signal with 
$m_\chi=8~\GEV$.
}
\label{heavyHiggs_ecal}
}

In the following, we take another approach based on jet substructure technique to identify the hadronic decays $h\to 2\chi\to 2j$. 

\vspace{.3cm}
{\bf Jet Substructure}

When reconstructing jets, one has to adopt an algorithm which iteratively merges protojets -- experimental objects such as calorimeter towers, clusters, or final state particles -- into jets. Therefore, jet recombination process in jet-finding algorithms, like CA and KT in our study, naturally preserves the characteristic substructure for a fat jet initiated from a boosted heavy particle decay. Recently,   the use of the jet internal structure has provided us a new window for reconstruction of a heavy particle which hadronically decays into a single jet, such as a boosted Higgs boson decay~\cite{VH_subjet} and a boosted top quark decay~\cite{top_tagger,ttH_subjet}. 
In our scenario,  since the pseudoscalar particle is highly boosted and  decays into a pair of massless gluons, we then expect that the jet from $\chi$
 decay contains rather symmetric and two hard substructures.  We follow the prescription in Ref.~\cite{VH_subjet} to find jet substructure as:

\begin{enumerate}
\item
Firstly, particles (excluding an isolated lepton) are clustered into jets with a radius $R$. 
For comparison, we use two clustering algorithms: CA and KT algorithms which have hierarchical structure for the clustering in angles and in relative transverse momenta, respectively. 
\item
For the two hardest jets, the substructure is searched for by undoing the last recombination step, namely  breaking each jet $j$ into two subjets, $j_1$ and $j_2$ such that $m_{j_1}>m_{j_2}$. 
\item
If there is a significant mass drop, $m_{j_1}<\mu m_j$ and the splitting is not too asymmetric, $y=\frac{\min(p^2_{Tj_1},p^2_{Tj_2})}{m^2_j}\Delta R^2_{j_1,j_2}\simeq\frac{\min(p_{Tj_1},p_{Tj_2})}{\max(p_{Tj_1},p_{Tj_2})}> y_{\rm cut}$, jet $j$ will be considered as a pseudoscalar particle neighborhood -- including the daughter gluons and QCD radiation -- and the finding process stops. 
\item
Otherwise, replace $j$ by $j_1$. If the ratio of $p_{Tj}$ to the original jet $p_T$ is larger than a parameter $\delta_p$, then go back to repeat from step 2; otherwise, the process stops.
\end{enumerate}

For this analysis, we take $R=0.5$ which is smaller than a typical value $R\sim 1.2-1.5$ in other literatures. This is because the pseudoscalar particle 
 $\chi$ is much lighter than its mother particle, the Higgs boson, and is highly boosted, therefore, the jet initiated from $\chi$ decay should be narrow. Other three parameters $\mu,~y_{\rm cut}$ and $\delta_p$ can be chosen independently. It was pointed out in \cite{VH_subjet} that the decay into a three equally energy shared configuration, i.e. $ggg$ configuration, will still trigger the mass drop condition if one takes $\mu\gtrsim 1/\sqrt{3}$. Here, we use $\mu=0.67$. The cut on $y$ is used to eliminate the asymmetric configurations from soft radiations that usually generate significant  jet mass and the parameter $\delta_p$ is introduced just to prevent the algorithm to go back too many steps. In this work, we take $y_{\rm cut}= (0.35)^2\sim 0.12$ and $\delta_p=0.2$.

In addition, we also expect that at the scale where two subjets are resolved, the KT distance, $d\equiv \min[p^2_{T}(subj_1),p^2_{T}(subj_2)]\Delta R^2_{subj_{12}}/R^2$, 
will be of the order $\mathcal{O}(m^2_\eta)$~\cite{Seymour, susy_subjet}, where $p_{T}(subj_{1,2})$ are the transverse momenta of the subjets $1$ and $2$, and $\Delta R^2_{subj_{12}}$ is the $\Delta R^2$ between subjets $1$ and $2$.
 The distributions of $\log(\sqrt{d})$ for signal and $Wjj$ background are shown in Figure \ref{ktdis}. We define the \textit{"subjet cut"} as the requirement that jet substructures  can be found  and  $0.75<\log(\sqrt{d})<2.0$ is satisfied for both of two hardest jets.

\FIGURE[!ht]{
\includegraphics[scale=0.6]{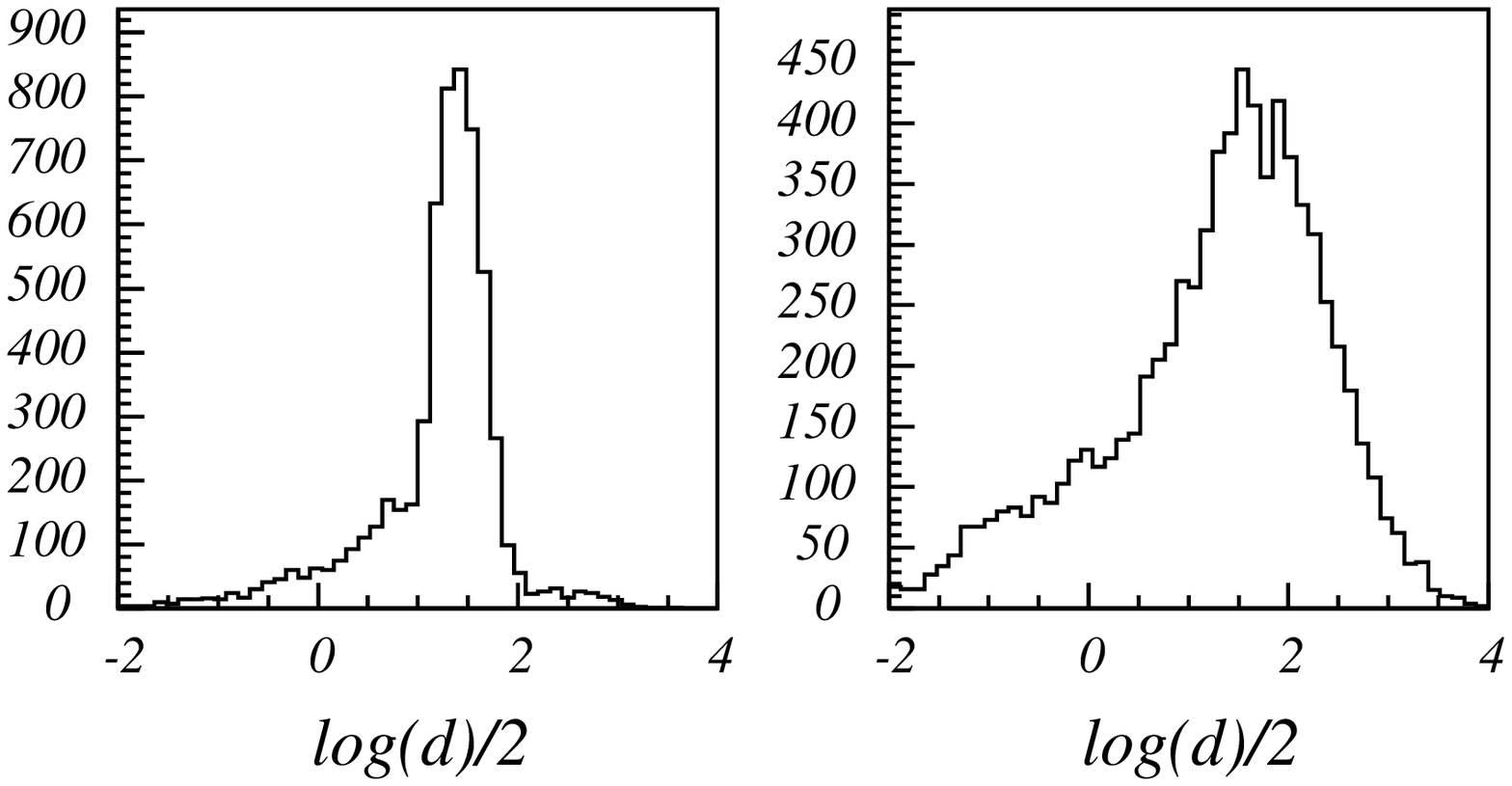}
\caption{Non-normalized distributions of $\log(\sqrt{d})$ for signal with 
$m_\chi=8~\GEV$ and $Wjj$ background.}
\label{ktdis}
}

\TABLE[!ht]{
\begin{tabular}{|l|r|r|r|r|r|r|r|}
\hline
Selection cut & $Wh$  & $Wjj$ & $WW$ & $t\bar t$ & $tq$ & $tW$ & $tbW$ \\
{}  & ($m_\chi=8~\GEV$) & & & & & & \\
\hline\hline
pre-selection cuts & 708 & 528,464 & 5,826 & 152,389 & 8,972 & 13,332 & 30,754 \\
\hline
$n_j\le 3$ & 578 & 393,372 & 3,796 & 6,612 & 6,704 & 3,410 & 1,314 \\
$m_j\le 12~\GEV$ & 435 & 175,254 & 2,739 & 30,949 & 3,728 & 3,596 & 6,288 \\
$p_{Tj_{1,2}}\ge 100,50~\GEV$ &  112 & 14,927 & 75 & 31,444 & 242 & 1,286 & 5,490 \\
\hline\hline
imposing all cuts (except & 43 & 513 & 5 & 81 & 14 & 28 & 22 \\
 subjet cut)&  &  &  &  &  &  & \\
after including subjet cut & 16 & 16 & 0 & 2 & 0 & 0 & 0\\
\hline
\end{tabular}
\caption{Number of expected signal and background events after cuts in the dijet invariant mass window 110~GeV$\le m_{jj}\le$ 130~GeV for $\mathcal{L}=30~{\rm fb}^{-1}$ at the LHC. The analysis is based on KT algorithm.}
\label{subj_kt_count}
}

The number of expected signal and background events after imposing cuts in the dijet invariant mass window 110~GeV$\le m_{jj}\le$ 130~GeV 
 are shown in Table \ref{subj_kt_count} for $\mathcal{L}=30~{\rm fb}^{-1}$ at the LHC. Again, the first row is the number of events after imposing pre-selection cuts, and  lower entries are numbers when cuts on number of jets, jet mass, and jet transverse momentum are separately applied. 

In this analysis, the largest background is $Wjj$ because of its very large cross section. However, $t\bar t$ background is also significant and difficult to eliminate. This is because $b$-quark from each top decay has an energy close to $m_h/2$ in the top rest frame and dijet invariant mass distribution from $t\bar t$ has a scale close to Higgs mass scale. By putting more stringent cut $n_j\le 3$
\footnote {The number of jets is sensitive to underlying events and effects of higher-order QCD calculations. If the number of events which contain more than two soft jets is significant, the stringent cut on number of jet, $n_j\le 3$, will largely reduce our signal. And also, the number of jets in $t\bar{t}$ background depends sensitively on $p_T$ and $|\eta|$. For example, when we impose $p_T> 20~\GEV$ and $|\eta|<3.5$ in jet definition, the $t\bar{t}$ events with $n_j\le3$ will increase by a factor of $3\sim4$, while the our signal and the main $Wjj$ background increase by factors of about $1.2$ and $1.3$, respectively.}
, $t\bar t$ background can be reduced by a factor of 20. The effect of jet mass and $p_T$ cuts are similar to the discussion in the previous analysis. However, the cut in this analysis is somehow looser so that many backgrounds can still survive. 

\FIGURE[!ht]{
\hskip -15pt
\includegraphics[scale=0.35]{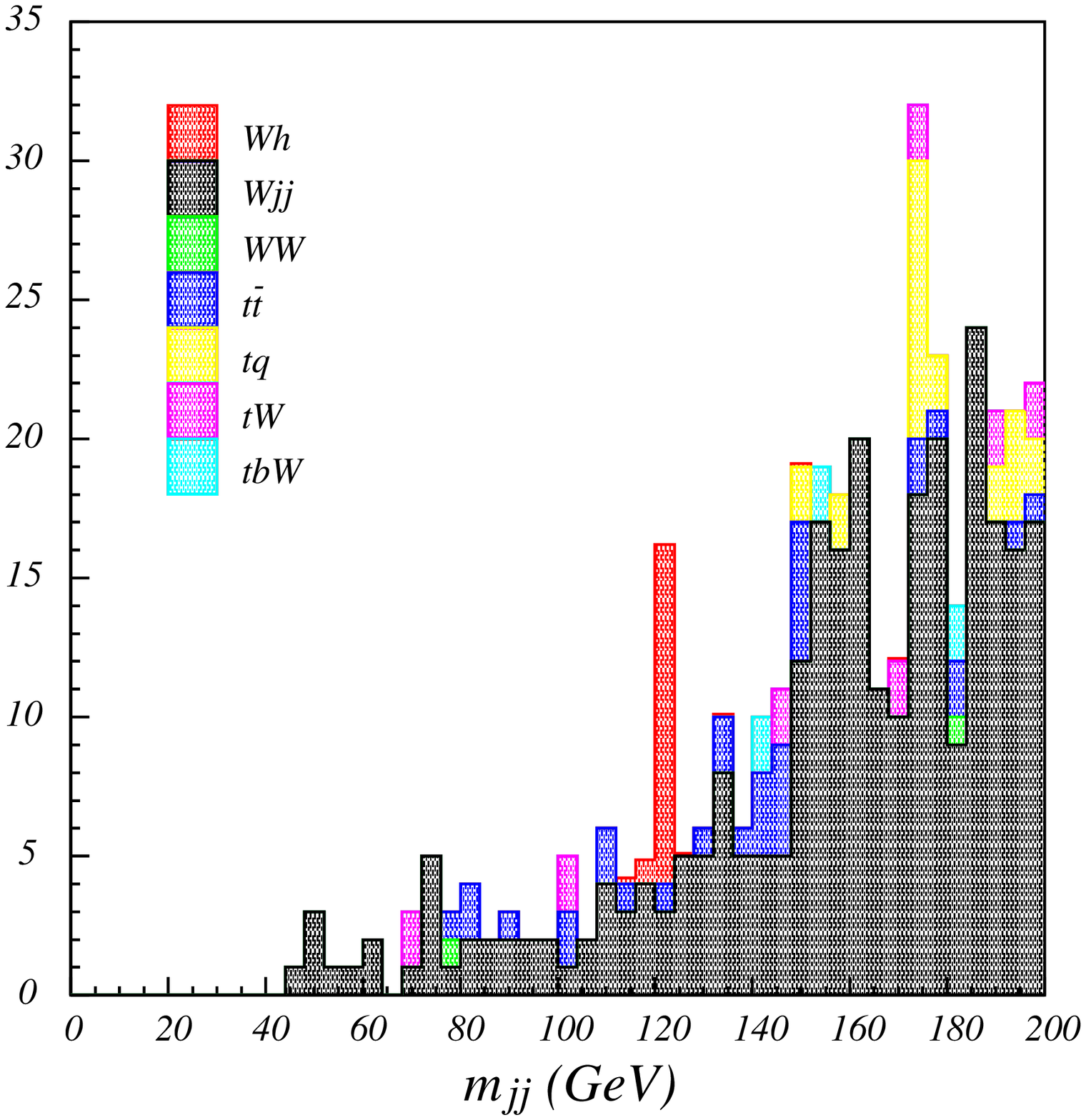}
\includegraphics[scale=0.35]{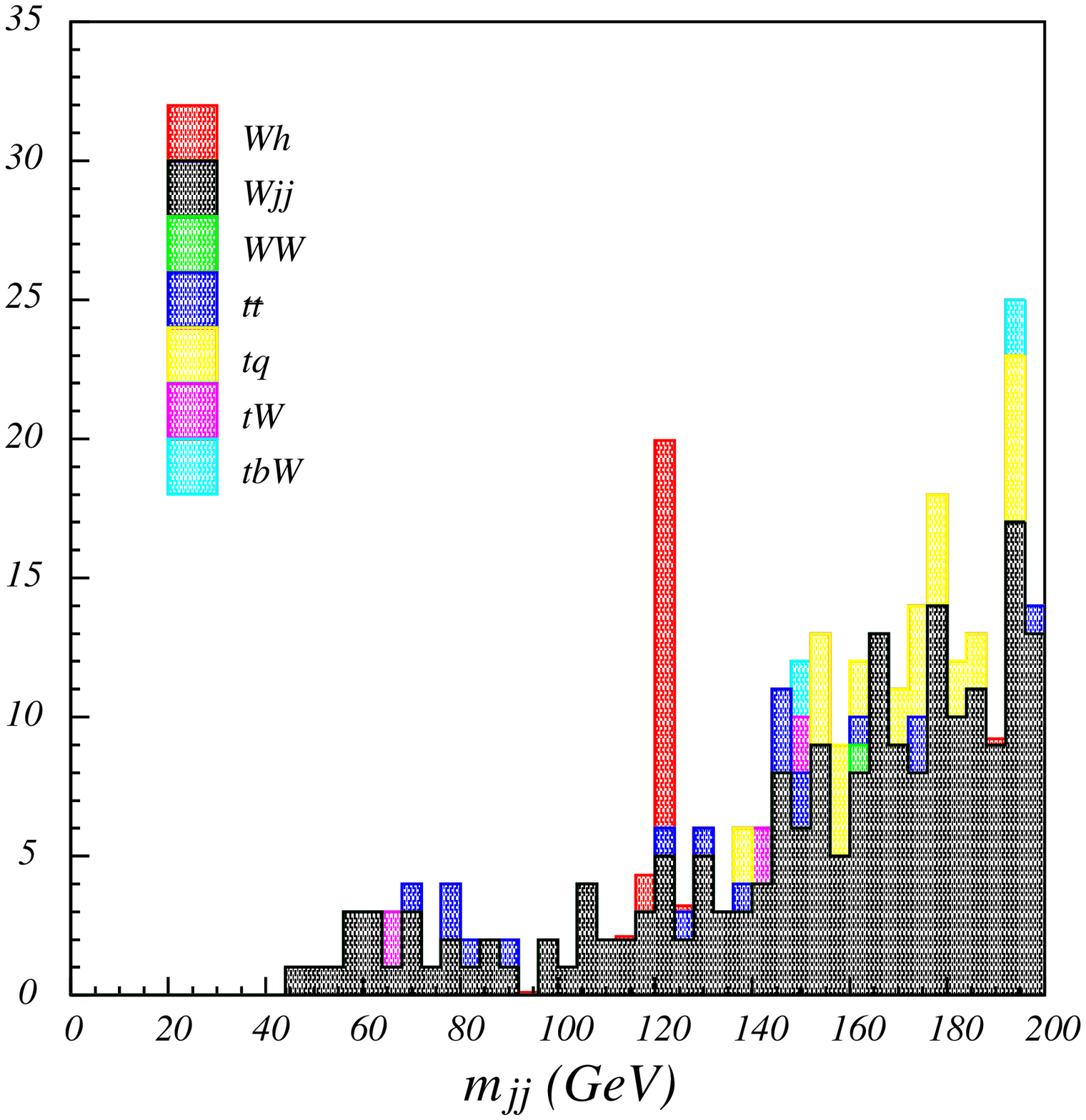}
\caption{Dijet invariant mass distributions for $pp\to Wh$ and 
$h\to 2\chi\to 2j$ $(m_\chi=8~\GEV)$ 
and SM backgrounds from jet substructure analysis using CA algorithm (left) and KT algorithm (right) for $\mathcal{L}=30~{\rm fb}^{-1}$ at the LHC.}
\label{hsubj_combine}
}

We show  in Table \ref{sigma_subj} the cross section of signal and statistical significance after imposing all cuts. It can be seen that, in the mass window between $110~\GEV$ and $130~\GEV$, one can reach a $2.65\sigma$ and $3.77\sigma$ statistical significance  using CA and KT jet-finding algorithms, respectively, with $30~{\rm fb}^{-1}$ luminosity at the LHC.
We also notice that the KT algorithm seems to perform better than the CA algorithm which is in contradiction of the conclusion from Ref. \cite{VH_subjet}.
However, it should be kept in mind that the angular splitting between two gluons from $\chi$ decay is small and the configuration is indeed different from one in Ref. \cite{VH_subjet}. In addition, the statistics is rather low and smearing effect is not taken into account in this analysis, so more realistic simulation should be performed before more decisive conclusion can be made.

\TABLE[!ht]{
\begin{tabular}{|l|c|c|}
\hline
Jet algorithm & $\sigma_S$ (fb) & $S/\sqrt{B}$\\
\hline
CA & 0.43 & 2.65 \\
KT & 0.53 & 3.77 \\
\hline
\end{tabular}
\caption{Signal cross section and statistical significance after all cuts in a dijet invariant mass window 110~GeV$\le m_{jj}\le$ 130~GeV from jet substructure analysis for $\mathcal{L}=30~{\rm fb}^{-1}$ at the LHC.}
\label{sigma_subj}
}

 Furthermore, the acceptances of the jet mass cut $m_j \le 12~\GEV$ are different for KT and CA algorithms. As we can see in the Table~\ref{subj_kt_count}, about $33\%$ and $20\%$ of our main background  $Wjj$ and $t\bar{t}$ events pass the jet mass cut, respectively, when using KT algorithm. However, we found that the acceptances increase up to about $40\%$ for $Wjj$ and $30\%$ for $t\bar{t}$ background events when CA algorithm is adopted\footnote{The acceptances of other cuts are similar for KT and CA algorithms. The event numbers of signal and total background are $13$ and $24$, respectively, after imposing all of the cuts using CA algorithm.}.

The dijet invariant mass distributions after including subjet cut are shown in Figure \ref{hsubj_combine}. The left and right plots are obtained by using CA and KT algorithms, respectively. 
We can see that the signal peak can be visible above backgrounds in both plots
and the invariant mass distribution for signal is very sharp since smearing is off as mentioned earlier.

\section{Discussions and Conclusions}

In this paper, we studied the signature and the discovery potential of an elusive Higgs boson which is produced in association with a $W$ boson at the LHC. The interesting phenomenology is that the Higgs boson decays dominantly into a pair of light pseudoscalar particles, $\chi$, which then decay into a pair of gluons if 
$m_\chi<2m_b$. The final state, in general,  contains at least four jets corresponding to four daughter gluons from the Higgs boson cascade  decay. However, due to its small mass, pseudoscalar becomes highly boosted; as a result, the two gluons from its decay move collinearly and appear as a single jet in the detector. Therefore, the signature of such Higgs boson is then two central narrow non-b-jets.

We proposed two methods to search for such elusive Higgs boson at the LHC: electromagnetic calorimeter-based and jet substructure analyses, depending on the mass of the $\chi$. We took two benchmark masses, 
$m_\chi =4~\GEV$ and $m_\chi =8~\GEV$,
 to demonstrate how these two methods work. Electromagnetic calorimeter-based variables are introduced for characterization of a very narrow jet from 
 $\chi$ decay. We showed that for $m_\chi =4~\GEV$ case, 
 most of jet energy is deposited around the jet core axis and the imposition of cut on electromagnetic calorimeter-based variables is very useful to suppress the background.  With $30~{\rm fb}^{-1}$ integrated luminosity, we can reach more than $7\sigma$ statistical significance at the LHC for a $120~\GEV$ Higgs boson. 
When $m_\chi$ is larger (but still $<2m_b$), e.g. $m_\chi = 8~\GEV$, 
activities inside a jet of signal spread out and distributions of electromagnetic calorimeter-based variables become broader and look like those from backgrounds. The cut on these variables is therefore not effective enough to dig the Higgs boson signal out of a huge $Wjj$ background. We then turned to use a jet substructure method, which recently has gained much attention as a powerful tool to search for hadronically decaying heavy particles. We demonstrated that the signal significance about $3.77\sigma$ can be obtained with  $30~{\rm fb}^{-1}$ integrated luminosity at the LHC when KT jet-finding algorithm is used.

Several comments are in order. We adopt the SM value for the Higgs production cross section and $100\%$ branching fraction for 
$h\to 2\chi$ and $\chi\to 2g$ 
decays, since our purpose is to show the feasibility of the discovery for such an elusive Higgs boson at the LHC. Our results can be easily rescaled for a specific model.
It should be also emphasized again that, in our analysis, we specifically forced $W$ boson to decay into electron for both signal and background processes. Therefore the total statistics should be double when the muon channel is considered as well.
 Lastly, the model of the underlying event (UE) in Herwig is known for the absence of multiple parton interaction component. We expect that our results are not crucially affected by its correction since we are considering a narrow jet regime in which the perturbative contribution is predominant.

\section*{Acknowledgments}

This work is supported in part by the World Premier International Center  
Initiative (WPI Program), MEXT, Japan and the Grant-in-Aid for Science  
Research, Japan Society for the Promotion of Science (for MN) and the Monbukagakusho (Japanese Government) Scholarship (for WS).

\appendix
\section{Appendix}

In this Appendix, we show distribution of electromagnetic calorimeter-based variables, $R_{\rm em}$ and $E_{\rm iso}$, for $h\to gg$, $h\to b\bar b$, $h\to u\bar u$ and $h\to\tau^+\tau^-$, respectively.
\newpage
\FIGURE[!ht]{
\hskip -5pt
\includegraphics[scale=0.45]{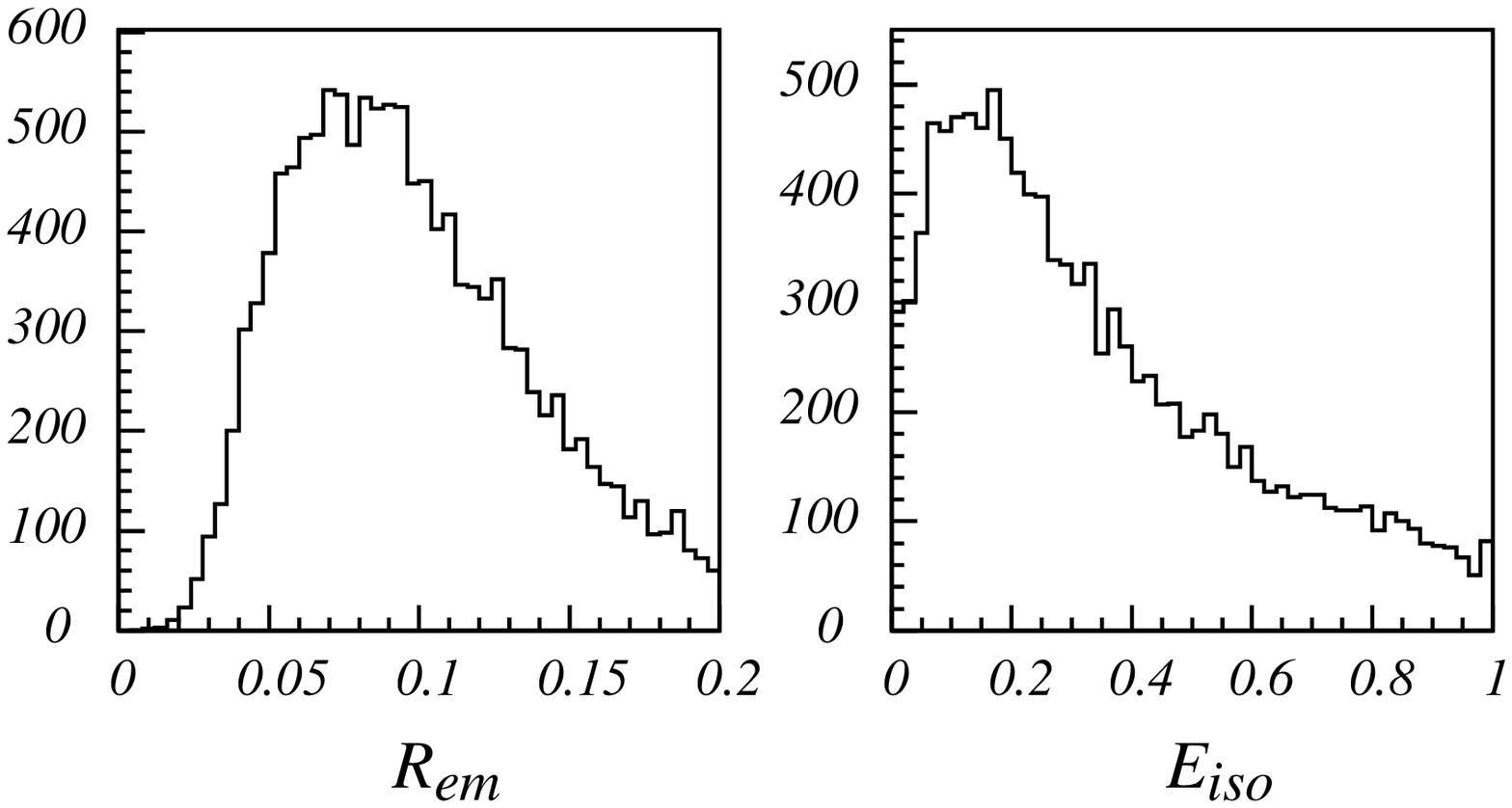}
\includegraphics[scale=0.45]{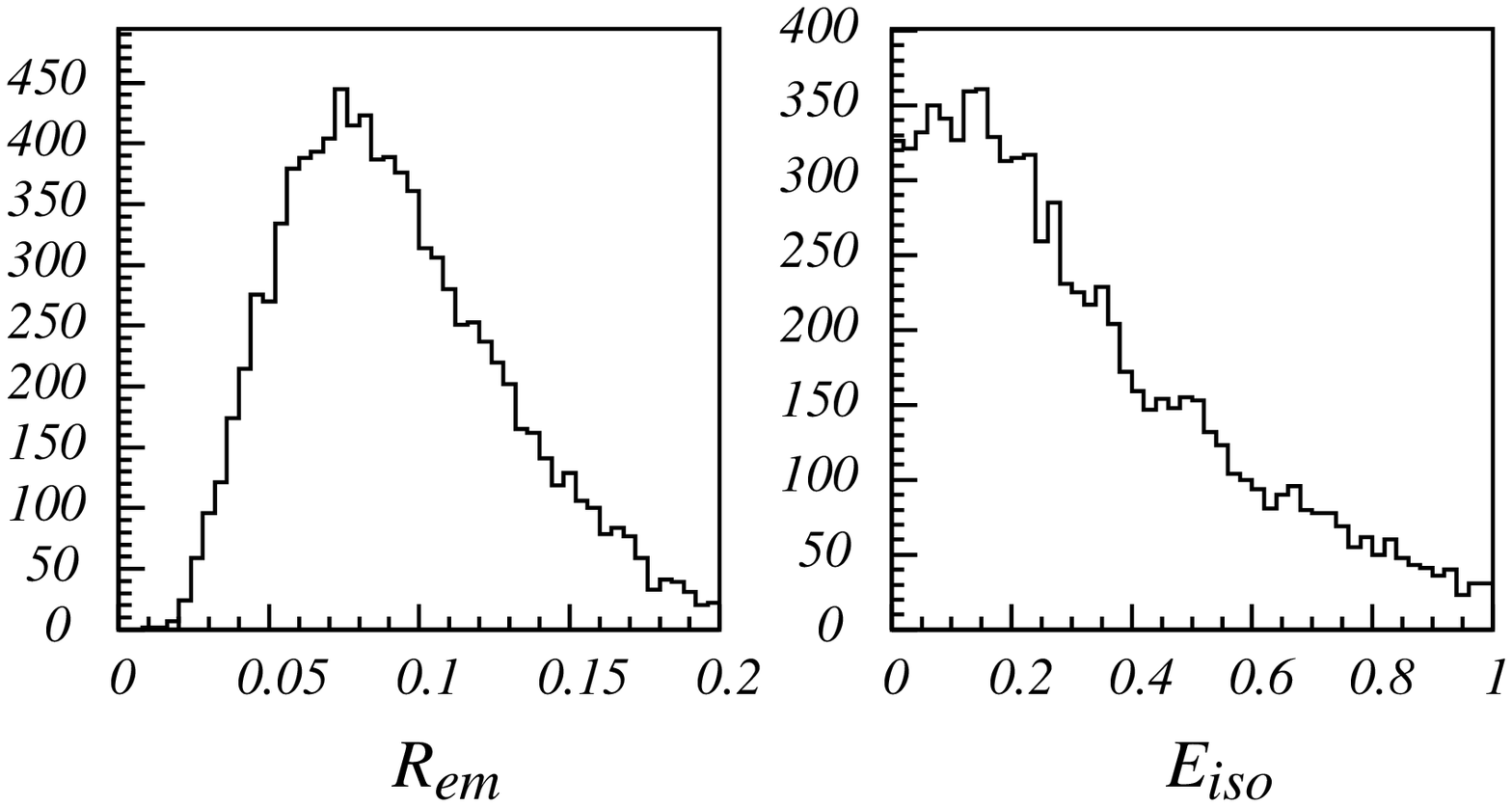}
\includegraphics[scale=0.45]{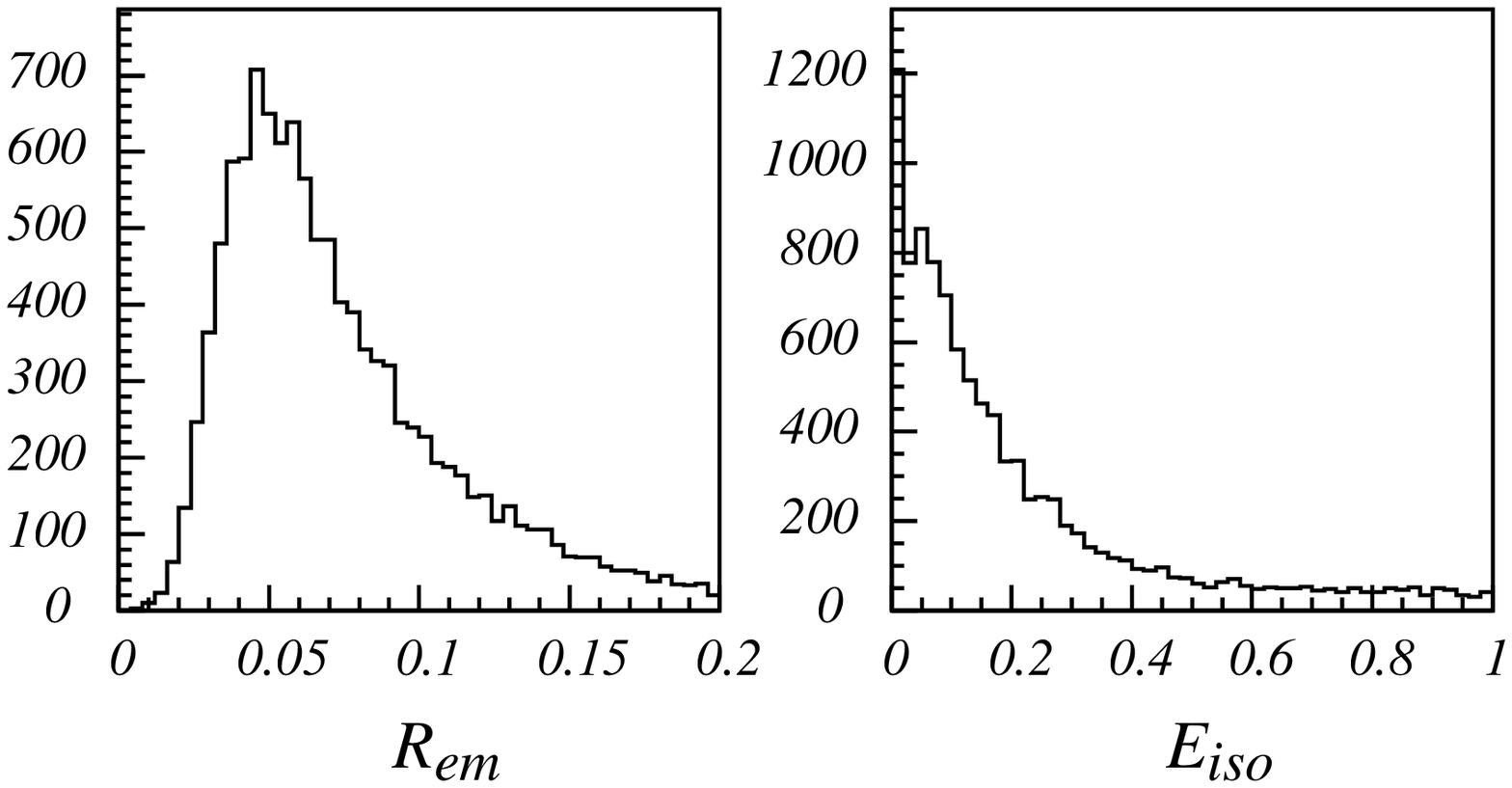}
\includegraphics[scale=0.45]{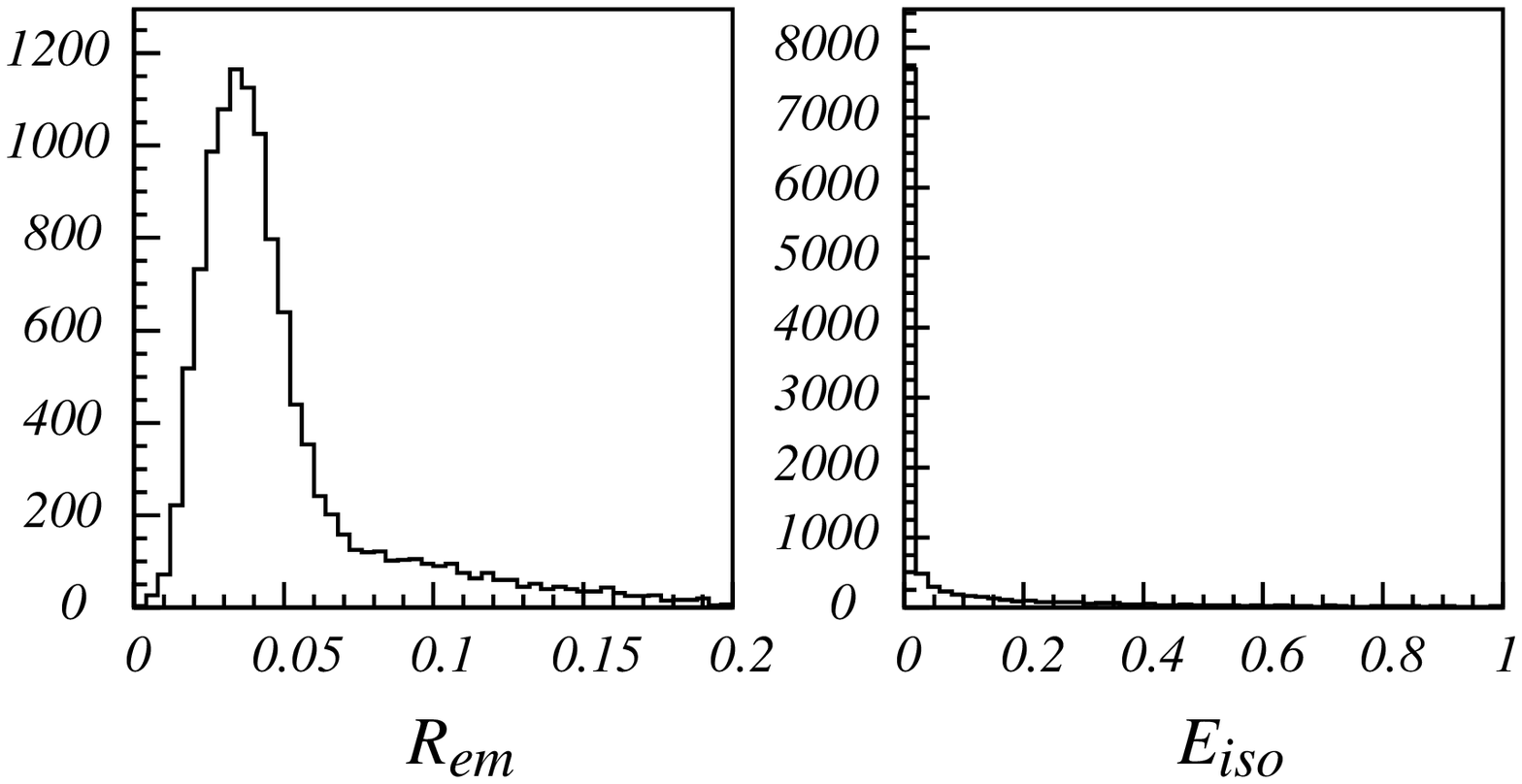}
\caption{Distributions of electromagnetic calorimeter-based variables, $R_{\rm em}$ and $E_{\rm iso}$,  for $h\to gg$ (upper left), $h\to b\bar b$ (upper right), $h\to u\bar u$ (lower left) and $h\to\tau^+\tau^-$ (lower right).}
\label{sm_ecal}
}


\end{document}